%% file: paper.tex
\documentclass[sigconf]{acmart}

\input{macros}

\renewcommand\footnotetextcopyrightpermission[1]{}
\setcopyright{none}
\acmConference[SAST 2026]{11st Brazilian Symposium on Systematic and Automated Software Testing}{September 8--11, 2026}{São Paulo, SP, Brazil}

\AtBeginDocument{
    
}

\graphicspath{{./}}

\begin{document}

\title[Evaluating Shaker for Flaky Test Detection in Python]{Evaluating Shaker for Flaky Test Detection in Python Projects}

\author{Gabriela Leal}
\authornote{These authors contributed equally to this work.}
\affiliation{
  \institution{Centro de Informática\\Universidade Federal de Pernambuco}
  \city{Recife}
  \country{Brazil}
}
\email{gabrielamfleal@gmail.com}

\author{Denini Silva}
\authornotemark[1]
\affiliation{%
  \institution{Universidade Federal Rural de Pernambuco}
  \city{Belo Jardim}
  \country{Brazil}
}
\email{denini.gabriel@ufrpe.br}

\author{Leopoldo Teixeira}
\affiliation{
  \institution{Centro de Informática\\Universidade Federal de Pernambuco}
  \city{Recife}
  \country{Brazil}
}
\email{lmt@cin.ufpe.br}

\renewcommand{\shortauthors}{Leal et al.}
\renewcommand{\shorttitle}{Evaluating Shaker for Flaky Test Detection in Python Projects}

\begin{abstract}
Flaky tests pass or fail non-deterministically on unchanged code, eroding 
trust in test suites and inflating the cost of every failure. Shaker detects 
them by injecting resource contention (CPU, memory, and I/O stress) to 
amplify non-determinism caused by concurrent execution, and was reported to 
detect 95\% of the
flaky tests in a Java and Android benchmark against 37.5\% for plain
re-execution (ReRun). We present the first empirical evaluation of Shaker for
Python. Drawing non-order-dependent flaky tests from the ground-truth dataset of Gruber et al., we
compare Shaker against a budget-matched ReRun baseline in a paired design,
giving both techniques the same number of test executions: Each of 137 tests is
run 100 times under each. As configured for
Java and Android, Shaker provides \emph{no} statistically significant detection
advantage over plain re-execution (37.2\% vs.\ 35.8\%; McNemar exact
$p = 0.84$). Two findings explain why. First, fewer than half of the
ground-truth flaky tests reproduce as flaky at all on independent hardware under
either technique, and most of the tests that fail to reproduce never diverge
once across 100 runs. Second, the tests that do reproduce are dominated by
flakiness from network interactions and randomness rather than the concurrency
Shaker targets. Beyond the tool, this exposes a broader hazard for the field:
reusing a flaky-test ground truth across execution environments silently
converts genuine flaky tests into apparent true negatives, deflating any tool's
measured recall. 
\end{abstract}


\frenchspacing
\maketitle

\input{sections/01_introduction}
\input{sections/03_illustrative_example}
\input{sections/02_background}
\input{sections/04_objects_and_setup}
\input{sections/06_evaluation}
\input{sections/07_threats}
\input{sections/08_related_work}
\input{sections/09_conclusions}

\section*{Artifact Availability}

Our replication package is publicly available at
\url{https://doi.org/10.5281/zenodo.22119340}. It contains the ground-truth flaky
tests, both result datasets, the Docker-based execution infrastructure, the
modified Shaker of Section~\ref{sec:procedure}, and the scripts that regenerate
every table and figure. Code is released under the BSD 2-Clause License and data
under CC BY 4.0, matching the dataset by
\citeauthor{gruber2021empirical}~\cite{gruber_2021_4450435} from which our
ground-truth files derive.

\section*{Acknowledgments}

This work was partially supported by INES.IA\footnote{\url{https://www.ines.org.br}}, CNPq grant 408817 /2024-0, and CNPq grant 310405/2025-4.

\bibliographystyle{ACM-Reference-Format}
\bibliography{biblio}

\end{document}

%% file: macros.tex
\usepackage{booktabs}     

\usepackage{graphicx}
\usepackage{subcaption}
\usepackage{placeins} 
\usepackage{float}     
\usepackage{tcolorbox}
\tcbuselibrary{skins,breakable,listings}
\usepackage{listings}
\usepackage{xcolor}  
\usepackage{array}
\lstdefinestyle{code}{
  basicstyle=\ttfamily\small,
  numbers=left, numberstyle=\scriptsize, numbersep=6pt,
  breaklines=true, breakatwhitespace=true,
  frame=single, framerule=0.4pt,
  tabsize=2, showstringspaces=false
}
\lstdefinestyle{cpp}{style=code, language=C++}
\lstdefinestyle{json}{style=code, language=}
\lstdefinestyle{txt}{style=code, language=}
\lstdefinestyle{cmd}{style=code, language=, numbers=none,
                     basicstyle=\ttfamily\footnotesize}

\newtcblisting{codecard}[2][]{%
  enhanced, breakable,
  colback=white, colframe=black!12,
  boxrule=0.6pt, arc=2mm, left=1mm, right=1mm, top=1mm, bottom=1mm,
  title={#2}, fonttitle=\bfseries\footnotesize,
  listing only, listing options={#1}
}

\usepackage{tabularx}

\usepackage{wrapfig}
\usepackage{multirow}
\usepackage{multicol}
\usepackage{enumitem}
\usepackage{listings}
\usepackage{xspace}
\usepackage{subcaption}
\usepackage{pgf-pie}
\usepackage{pifont}

\usepackage{tikz}
\tikzset{>=latex}
\usetikzlibrary{calc}
\usetikzlibrary{shapes.geometric}
\usetikzlibrary{decorations.pathreplacing}
\usetikzlibrary{positioning}
\usetikzlibrary{arrows.meta}
\usetikzlibrary{fit}
\usetikzlibrary{backgrounds}
\newcommand*\circled[1]{\tikz[baseline=(char.base)]{
            \node[shape=circle,draw,fill=black,text=white,inner
            sep=0.4pt] (char) {\small #1};}}

\newcommand{\cmark}{\color{blue}{\ding{51}}}
\newcommand{\xmark}{\color{red}{\ding{55}}}

\usepackage[framemethod=tikz]{mdframed}
\mdfdefinestyle{mpdframe}{
   frametitlebackgroundcolor   =black!15,
    frametitlerule              =true,
    roundcorner                 =3pt,
    middlelinewidth             =1pt,
    skipabove                   =\topskip,
    skipbelow                   =\topskip,
    innermargin                 =0.1cm, 
    outermargin                 =0.1cm,
    innerleftmargin             =0.1cm,
    innerrightmargin            =0cm,
    innertopmargin              =0.1cm,
    innerbottommargin           =0.1cm,
    align=center   
}

\newcommand{\DefMacro}[2]{%
   \expandafter\newcommand\csname rmk-#1\endcsname{#2}%
}
\newcommand{\UseMacro}[1]{\csname rmk-#1\endcsname}

\newcommand{\Comment}[1]{}

\newcommand{\Space}[1]{}

\newcommand{\MyPara}[1]{\vspace{1pt}\noindent\textbf{#1}.}

\newcommand{\Code}[1]{{\small\ifmmode{\texttt{#1}}\else$\texttt{#1}$\fi}}
\newcommand{\CodeIn}[1]{{\small\ifmmode{\mathtt{#1}}\else$\mathtt{#1}$\fi}}
\newcommand{\ColorBack}[1]{%
  \begingroup \setlength{\fboxsep}{0pt}
}

\definecolor{gray}{RGB}{211,211,211}

\newcommand{\EQ}[1]{RQ#1\xspace}

\newcommand{\todo}[2][normal]{%
  \ifdefined\showtodos
    \ifstrequal{#1}{high}{\sethlcolor{red}}{}%
    \ifstrequal{#1}{normal}{\sethlcolor{yellow}}{}%
    \ifstrequal{#1}{low}{\sethlcolor{green}}{}%
    {\hl{\textbf{TODO:} #2}}%
    \marginpar{\textcolor{red}{\textbf{TODO}}}%
  \fi
}

\newif\ifshowtodos
\showtodosfalse  

\DefMacro{eq-single-performance}{\EQ{1}}
\DefMacro{eq-single-safety}{\EQ{2}}
\DefMacro{eq-evolution-stability}{\EQ{3}}
\DefMacro{eq-evolution-performance}{\EQ{4}}
\DefMacro{eq-evolution-safety}{\EQ{5}}
\DefMacro{Category}{Category}
\DefMacro{PolicyType}{Policy Type}
\DefMacro{Description}{Description}

\definecolor{SubtleColor}{rgb}{0,0,.50}

\definecolor{DefensiveColor}{RGB}{65,105,225}      
\definecolor{PortableColor}{RGB}{34,139,34}        
\definecolor{NormColor}{RGB}{205,92,92}            
\definecolor{EnvColor}{RGB}{138,43,226}            

\definecolor{UniqueSymptomColor}{RGB}{200,230,255}     
\definecolor{ContextSymptomColor}{RGB}{255,250,205}    

\newcommand{\ShakerRQOne}{At an equal execution budget, does Shaker's resource-stress injection detect more flaky tests in Python projects than plain re-execution (ReRun)?}
\newcommand{\ShakerRQTwo}{To what extent do known-flaky Python tests reproduce as flaky under repeated execution on independent infrastructure?}
\newcommand{\ShakerRQThree}{Which test and project characteristics distinguish the flaky tests that reproduce (and are detected) from those that do not?}

\newtcolorbox{finding}{%
  enhanced, breakable,
  colback=black!3, colframe=black!55,
  boxrule=0.5pt, arc=1mm,
  left=2mm, right=2mm, top=1mm, bottom=1mm}

%% file: sections/01_introduction.tex
\section{Introduction}
\label{sec:introduction}

A test carries an implicit contract: For a given version of the code, it should
deterministically signal the presence or absence of a defect. When that contract breaks
down, so does developer trust in the test suite. A \textit{flaky test} is a test that
produces non-deterministic pass or fail verdicts without any change to the code under
test, breaking this contract~\cite{luo2014empirical, parry2021survey}.

The consequences compound over time. When a test fails non-deterministically, it is
initially unclear whether a real bug has been introduced or a flaky test has fired, forcing
developers to investigate and rerun tests instead of trusting the
signal~\cite{eck2019understanding}. A test suite known to contain flaky tests eventually
becomes a liability: Developers learn to dismiss failures, and the reliability of the whole
suite is undermined.

Flaky tests are costly and widespread. At Google, 16\% of tests exhibit
flakiness~\cite{Google,memon2017taming}, requiring the company to maintain a dedicated team for
monitoring and support. 
\citeauthor{eck2019understanding}~surveyed 121 professional developers and found that
reproducing flakiness and diagnosing its root cause are the most difficult challenges,
leading many teams to simply rerun failing tests, a workaround that hides the symptom
without addressing its cause~\cite{eck2019understanding}. This naive strategy, 
known as \textit{ReRun}, is expensive in practice: \citeauthor{gruber2021empirical}~found 
that approximately 170 re-executions are needed per test to achieve 95\% confidence that a 
passing test is not flaky~\cite{gruber2021empirical}.

Several techniques have been proposed to detect flaky tests more efficiently than ReRun.
DeFlaker~\cite{bell2018deflaker} monitors code coverage changes between test runs,
flagging a failure as potentially flaky when it touches no code changed since the last
passing run; on 96 Java projects it detected 1,874 flaky tests among 4,846 failures with
only a 1.5\% false-alarm rate. iDFlakies~\cite{lam2019idflakies} detects
order-dependent tests by systematically reordering test suites and tracking inconsistent
verdicts. Shaker~\cite{silva2020shake} takes a complementary approach: Rather than
changing execution order or monitoring coverage, it \textit{injects} resource contention
(CPU, memory, and I/O stress) directly into the execution environment, amplifying
concurrency-induced non-determinism so that flaky tests fail more consistently and
are easier to identify. On a benchmark of 75 flaky tests from 11 Android applications,
Shaker detected 95\% of them against 37.5\% for a budget-matched ReRun, and surfaced 61
flaky tests that 50 re-executions missed~\cite{silva2020shake, 9678918}.

Shaker has nonetheless never been evaluated for Python, one of the most widely used
programming languages~\cite{ieee}, whose large open-source testing community centers on the
Pytest framework. Python's runtime characteristics differ substantially from Java's: The
Global Interpreter Lock (GIL) constrains true thread parallelism, the garbage collector and
memory model operate differently, and the concurrency primitives used by Python developers
follow different idioms. These differences raise a natural question: Do the stress
configurations that made Shaker effective for Java and Android transfer to Python, or does
Python's distinct execution environment require dedicated calibration?

To answer this question, we present the first empirical evaluation of Shaker for
Python. We draw flaky tests from the non-order-dependent portion of the large-scale
ground-truth dataset by Gruber et al.~\cite{gruber2021empirical} (876,186 test cases from
22,352 open-source PyPI projects), without \emph{further} pre-filtering by root cause so that
the sample reflects how the tool is used in practice, where a test's cause is unknown beforehand,
and evaluate Shaker against a budget-matched re-execution baseline on the 137
ground-truth flaky tests we could execute under both techniques. Matching the budget
means both techniques observe each test the same number of times, so any difference
is attributable to the stress mechanism rather than to the number of observations.
We address three research questions:

\begin{description}
  \item[\textbf{RQ1:}] \ShakerRQOne
  \item[\textbf{RQ2:}] \ShakerRQTwo
  \item[\textbf{RQ3:}] \ShakerRQThree
\end{description}

Our results show that, with the stress configuration inherited from Java and
Android, Shaker provides no statistically significant detection advantage over
plain re-execution at an equal execution budget. Two deeper reasons account for
this. Fewer than half of the ground-truth flaky tests reproduce as flaky at all
on independent hardware under either technique, and the tests that do reproduce
are dominated by flakiness from network interactions and randomness in smaller
projects, rather than the concurrency Shaker targets. The strong advantage
reported for Java and Android therefore does not transfer to our Python setting.


The main contributions of this paper are:

\begin{enumerate}
  \item The first empirical evaluation of Shaker for Python, showing that at an equal
        execution budget its stress injection detects no more ground-truth flaky tests
        than plain re-execution, in sharp contrast to the large advantage reported for
        Java and Android~\cite{silva2020shake};
  \item A cross-environment reproducibility analysis covering every non-order-dependent
        flakiness category in Gruber et al.'s ground truth, establishing that fewer than half
        reproduce on independent hardware and that this gap, rather than the detection 
        mechanism, is the dominant limiting factor, corroborating and generalizing a 
        similar gap previously observed for order-dependent Python tests
        specifically~\cite{wang2022ipflakies} into a cautionary result for any study 
        reusing a flaky-test ground truth collected elsewhere;
  \item A characterization of which flakiness categories the stress mechanism can and
        cannot reach in Python, which explains the null result;
  \item Two open-source enhancements to Shaker, that enable single-test execution and
        heterogeneous dependency-format support, which made this evaluation
        possible and are reusable for future Python flaky-test studies.
\end{enumerate}

%% file: sections/03_illustrative_example.tex
\section{Motivating Example}
\label{sec:motivation}

Shaker is built on a specific hypothesis: that many flaky tests fail because of \emph{resource- and
timing-sensitive} behavior that a lightly loaded machine hides, and that deliberately stressing the
machine makes such failures surface. Consider \texttt{test\_timer\_run} from the \texttt{cvbase}
project~\cite{cvbase_repo} (Listing~\ref{lst:cvbase}), one of the flaky tests in our sample. It starts a timer, sleeps
for one second, and asserts that the measured elapsed time is within 10 milliseconds of one second.

\begin{lstlisting}[language=Python, caption={A timing-sensitive flaky test from \texttt{cvbase}~\cite{cvbase_repo}.}, label={lst:cvbase}]
def test_timer_run():
    timer = cvb.Timer()
    time.sleep(1)
    assert abs(timer.since_start() - 1) < 1e-2 # within 10ms of 1s
    time.sleep(1)
    assert abs(timer.since_last_check() - 1) < 1e-2
    assert abs(timer.since_start() - 2) < 1e-2
\end{lstlisting}

On an idle machine the sleep is accurate and the assertion holds every time, so plain re-execution
never observes a failure: Across 100 unstressed runs the test passed 100 times. Under Shaker,
however, the injected CPU contention delays the process past the 10-millisecond tolerance often
enough that the test failed 6 of 100 times, enough to flag it as flaky. This is exactly the
amplification effect that made Shaker effective on Java and Android~\cite{silva2020shake}: Stress
converts a rarely triggered timing failure into an observable one.

The question this paper investigates is whether that effect generalizes to Python's flaky tests as a
whole. As Section~\ref{sec:results} shows, tests such as \texttt{test\_timer\_run}, where stress has a
mechanism to exploit, turn out to be the exception rather than the rule: Most Python flakiness in
our sample stems from causes such as randomness and network behavior that stress cannot influence,
and across the full sample Shaker detects no more flaky tests than plain re-execution.

%% file: sections/02_background.tex
\section{Background}

This section defines flaky tests and their costs, focuses on what is known about
flakiness in Python, and introduces Pytest and Shaker, the technique we
evaluate.

\subsection{Flaky Tests}

A \textit{flaky test} is a test that produces non-deterministic outcomes,
passing or failing when executed repeatedly on the same version of the code
without any changes to the test or the system under
test~\cite{luo2014empirical,parry2021survey}. Because a failing test is meant to
signal a genuine regression, flakiness erodes developer confidence in the entire
test suite~\cite{eck2019understanding}.

\MyPara{Root causes} \citeauthor{luo2014empirical}~\cite{luo2014empirical} established the first
empirical taxonomy of flaky tests by analyzing 201 commits addressing
flakiness in 51 open-source Apache and GitHub projects written in Java. They
identified ten root-cause categories: asynchronous waits, concurrency, test
order dependency, resource leaks, network communication, time, I/O,
randomness, floating-point arithmetic, and unordered collections, with
asynchronous waits, concurrency, and order dependency being the most
prevalent. The survey by \citeauthor{parry2021survey}~\cite{parry2021survey}, 
which synthesizes 76 papers across four dimensions (causes, costs, detection 
strategies, and mitigation), confirms this taxonomy as the community standard, 
while noting that the relative prevalence of each category varies significantly 
across programming ecosystems and test types.

\MyPara{Impact on developers and CI/CD} Flaky tests impose concrete costs on
development teams and continuous integration pipelines. 
\citeauthor{eck2019understanding}~\cite{eck2019understanding} surveyed 121 
professional developers about
their experience with flaky tests and found that \textit{reproducing}
flakiness and diagnosing its root cause are the most challenging aspects. In
the absence of dedicated tooling, developers dismiss flaky failures by
rerunning the test suite, a workaround that masks the underlying problem
without resolving it. A large-scale longitudinal study by Lam
et al.~\cite{lam2020longitudinal}, tracking flaky tests across 26 open-source
projects over six years, found that 85\% of flaky tests are flaky at the time
they are first introduced or directly modified, evidence that proactive
detection at test introduction is more effective than retrospective analysis.
At industrial scale, Google has reported that 16\% of its tests exhibit
flakiness and that flaky failures pervade its continuous-integration pipeline,
requiring dedicated tooling and a support team~\cite{Google,memon2017taming}.

\MyPara{Flakiness in Python} \citeauthor{gruber2021empirical}~\cite{gruber2021empirical}
conducted the first large-scale empirical study of flaky tests in Python,
analyzing 876,186 test cases from 22,352 open-source PyPI projects and
identifying 7,571 flaky tests. Their findings reveal a flakiness profile that
differs markedly from Java's: \textit{Order dependency} is the dominant cause in
Python (59\% of flaky tests), followed by infrastructure issues (28\%), while
concurrency, the primary target of Shaker, accounts for a much smaller
fraction. This difference is partly explained by Python's Global
Interpreter Lock (GIL), which constrains true thread-level parallelism and
reduces the prevalence of data races relative to Java. The gap between
Java-tuned tools and Python's flakiness profile is a central motivation for
our study.

\MyPara{Detection challenge} Detecting flakiness inherently requires observing
non-deterministic verdict variability across multiple executions of the same
test. \citeauthor{gruber2021empirical}~\cite{gruber2021empirical} estimated that approximately
170~re-executions are needed to achieve 95\% confidence that a passing test is
not flaky, a budget that makes naive rerun approaches expensive at scale.
This cost has motivated two complementary research streams. \textit{Dynamic
approaches} reduce the required execution budget by focusing test effort: For
instance, DeFlaker~\cite{bell2018deflaker} exploits coverage change to prioritize
candidate flaky tests, iDFlakies~\cite{lam2019idflakies} applies
order-permutation strategies to surface order-dependent tests, and
Shaker~\cite{silva2020shake,9678918} amplifies concurrency-induced
non-determinism via controlled resource stress. \textit{Static and predictive
approaches} instead avoid extra executions altogether;
FlakeFlagger~\cite{alshammari2021flakeflagger}, for example, predicts test
flakiness from behavioral features. Section~\ref{sec:related_work} surveys this
field in detail.

\subsection{Pytest}

Pytest~\cite{Pytest} is a well-established open-source testing framework for Python projects. In this study, Shaker's Python support is built on top of Pytest, using the \texttt{--junitxml} flag to produce structured test output reports that Shaker processes to determine test verdicts.

\subsection{Shaker}

Shaker is a tool designed to detect flakiness in codebases by inserting noise and load into the test execution environment~\cite{silva2020shake}. The key idea is that, despite the additional cost of running tests under stress, the tool can detect flaky tests faster than simply re-executing them (the ReRun method). This claim was established on Java and Android projects~\cite{silva2020shake,9678918}; whether it also holds for Python is precisely the question this work evaluates.

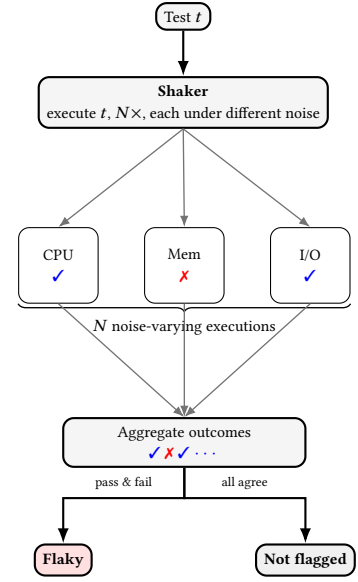
\begin{figure}
\centering
\begin{tikzpicture}[
  font=\scriptsize,
  >=Latex,
  box/.style={rounded corners, draw, thick, align=center, fill=black!4, inner sep=3pt},
  run/.style={rounded corners, draw, align=center, fill=white, inner sep=2pt,
              minimum width=10.5mm, minimum height=10mm},
  flow/.style={-{Latex[length=2mm]}, line width=0.8pt},
  fan/.style={-{Latex[length=1.5mm]}, line width=0.5pt, black!55},
  brace/.style={decorate, decoration={brace, amplitude=3pt, mirror}, line width=0.5pt},
  chip/.style={rounded corners, draw, thick, align=center, inner sep=3pt, font=\scriptsize\bfseries},
]

\node[box] (test) {Test $t$};
\node[box, below=6mm of test, minimum width=30mm]
  (shaker) {\textbf{Shaker}\\[1pt] execute $t$, $N{\times}$, each under different noise};
\draw[flow] (test) -- (shaker);

\node[run, below=13mm of shaker, xshift=-16.5mm] (r1) {CPU\\[1pt]\cmark};
\node[run, right=6mm of r1] (r2) {Mem\\[1pt]\xmark};
\node[run, right=6mm of r2] (r3) {I/O\\[1pt]\cmark};
\foreach \r in {r1,r2,r3}
  \draw[fan] (shaker.south) -- (\r.north);
\draw[brace] (r1.south west) -- (r3.south east)
  node[midway, yshift=-9pt, font=\scriptsize] {$N$ noise-varying executions};

\node[box, below=15mm of r2, minimum width=32mm] (agg)
  {Aggregate outcomes\\[1pt] {\scriptsize\cmark~\xmark~\cmark~$\cdots$}};
\foreach \r in {r1,r2,r3}
  \draw[fan] (\r.south) -- (agg.north);

\node[chip, below=10mm of agg, xshift=-16mm, fill=red!12] (flaky) {Flaky};
\node[chip, below=10mm of agg, xshift=16mm, fill=black!6] (notflaky) {Not flagged};
\draw[flow] (agg.south) -- ++(0,-4mm) -| node[pos=0.25, above, font=\tiny]{pass \& fail} (flaky.north);
\draw[flow] (agg.south) -- ++(0,-4mm) -| node[pos=0.25, above, font=\tiny]{all agree} (notflaky.north);
\end{tikzpicture}
\caption{Shaker's detection workflow. A target test is executed $N$ times, each run under a
different injected noise configuration; the resulting pass/fail verdicts are aggregated into an
outcome history, and the test is flagged as flaky when that history mixes passes and fails.}
\label{fig:shaker-workflow}
\end{figure}

Figure~\ref{fig:shaker-workflow} summarizes how Shaker turns a single test into a flakiness verdict. Given a target test, Shaker executes it $N$ times, and each execution is placed under a different noise configuration that perturbs the environment (e.g., CPU, memory, I/O, or scheduling interference), rather than repeating the exact same conditions every time. Every run produces an independent pass or fail verdict, and Shaker aggregates these verdicts into a per-test outcome history. If that history contains at least one pass and at least one fail, the test is flagged as flaky; if all $N$ runs agree, no flakiness is reported for the current budget. It is this diversity of injected noise across repeated executions, rather than sheer repetition alone, that Shaker relies on to raise the chance of exposing non-deterministic behavior.

To generate stress in the test execution environment, Shaker uses the stress-ng package~\cite{stressng}, which can configure stress loads on the CPU, virtual memory, disk, and other resources.

Shaker supports two modes of operation: as a GitHub Action that can be integrated into any GitHub project, and through a command-line interface (CLI)~\cite{9678918}. In this work we use the CLI mode, as it is simpler to automate. For test execution, Shaker supports Maven~\cite{maven} for Java projects and Pytest for Python projects. Listing~\ref{lst:shaker-format} gives the general format of the original Shaker command.

\begin{lstlisting}[style=txt, float=h, caption={Shaker command format.}, label={lst:shaker-format}]
shaker.py [-e EXTRA_ARGUMENTS] [-o OUTPUT_FOLDER]
[-sr STRESS_RUNS] [-nsr NO_STRESS_RUNS] {pytest,maven} directory
\end{lstlisting}

The two arguments that govern detection are \texttt{-sr} and \texttt{-nsr}, which set how many
stressed and unstressed runs the tool performs. They are not symmetric: Specifying $N$ stress runs
makes Shaker execute the test $N \times 4$ times, once under each of its four stress-ng
configurations, whereas $N$ no-stress runs yield $N$ plain executions. If the test fails in any
execution, the result is included in the final report. Section~\ref{sec:procedure} sets exactly
these two arguments to place Shaker and the re-execution baseline on an equal execution budget.

%% file: sections/04_objects_and_setup.tex
\section{Study Design}
\label{sec:design}

We designed our study to test whether Shaker's resource-stress strategy transfers
from Java and Android to Python.

\begin{figure*}
\centering
\begin{tikzpicture}[
  font=\small, >=Latex,
  stage/.style={rounded corners, draw, thick, align=center, text width=25mm,
                minimum height=18mm, inner sep=5pt, fill=black!4},
  rq/.style   ={rounded corners, draw, align=left, text width=34mm,
                minimum height=11mm, inner sep=4pt, fill=white, font=\scriptsize},
  panel/.style={rounded corners, draw, thick, dashed, fill=black!2, inner sep=6pt},
  flow/.style ={-{Latex[length=2.8mm]}, line width=1pt},
  fan/.style  ={-{Latex[length=2mm]}, line width=0.7pt, black!55},
  io/.style   ={font=\scriptsize, text=black!55, midway, above=2.5pt},
  note/.style ={font=\scriptsize, text=black!55},
  badge/.style={circle, fill=black, text=white, font=\scriptsize\bfseries, inner sep=1.3pt}
]
\node[stage] (s1)
  {\textbf{Object selection}\\[3pt]{\scriptsize Flaky tests from Gruber et al.~\cite{gruber2021empirical}.}};
\node[stage, right=11mm of s1] (s2)
  {\textbf{Detection}\\[3pt]{\scriptsize Repeated execution under resource stress and plain re-execution, at an equal budget}};
\node[stage, right=11mm of s2] (s3)
  {\textbf{Outcome classification}\\[3pt]{\scriptsize Each technique \emph{surfaces} the known flakiness (a pass \emph{and} a fail) or \emph{misses} it}};

\node[rq, anchor=west] (rq1) at ($(s3.east)+(16mm,15mm)$)
  {\textbf{RQ1: Baseline.} Shaker vs.\ ReRun at an equal execution budget};
\node[rq, anchor=west] (rq2) at ($(s3.east)+(16mm,0mm)$)
  {\textbf{RQ2: Reproducibility.} How many known-flaky tests reproduce at all};
\node[rq, anchor=west] (rq3) at ($(s3.east)+(16mm,-15mm)$)
  {\textbf{RQ3: Traits.} What distinguishes reproduced from missed tests};
\begin{scope}[on background layer]
\node[panel, fit=(rq1)(rq2)(rq3)] (s4) {};
\end{scope}
\node[note, text=black, anchor=south west, font=\scriptsize\bfseries]
  at ([xshift=4.5mm,yshift=1pt]s4.north west) {Empirical analysis};

\draw[flow] (s1) -- node[io]{tests} (s2);
\draw[flow] (s2) -- node[io]{verdicts}  (s3);
\coordinate (hub) at ($(s3.east)!0.5!(s4.west)$);
\draw[flow] (s3.east) -- node[io]{labels} (hub);
\draw[fan] (hub) |- (rq1.west);
\draw[fan] (hub) -- (rq2.west);
\draw[fan] (hub) |- (rq3.west);

\foreach \i/\n in {s1/1,s2/2,s3/3,s4/4}
  \node[badge] at (\i.north west) {\n};
\end{tikzpicture}
\caption{Overview of the study. Stages 1--3 form the detection pipeline and produce, for every
test and technique, a label recording whether the technique surfaced its known flakiness; stage 4
interprets these labels through three research questions (RQ1--RQ3).}
\label{fig:overview}
\end{figure*}
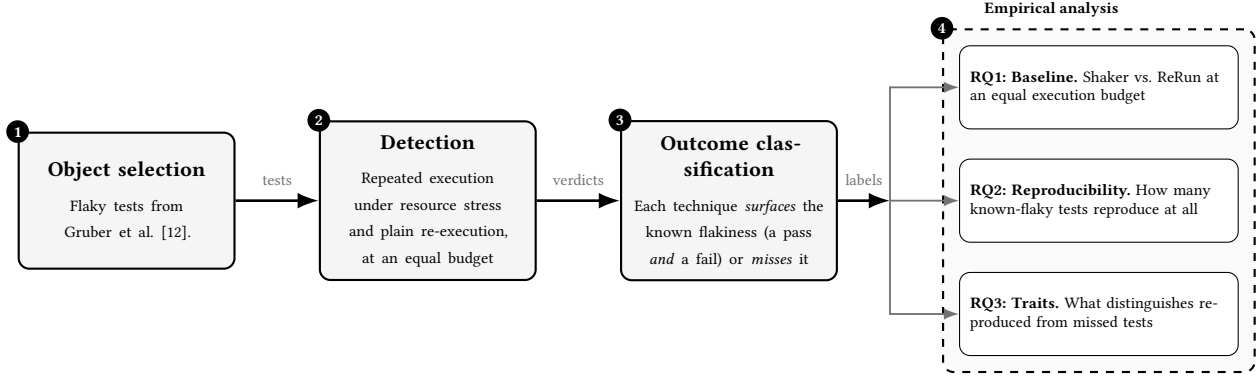

\MyPara{Workflow overview} As Figure~\ref{fig:overview} shows, the study proceeds in four stages.
\circled{1}~\emph{Object selection} draws ground-truth flaky tests from Gruber
et al.~\cite{gruber2021empirical}. \circled{2}~\emph{Detection} executes each test repeatedly under
Shaker's resource stress and, at a matched execution budget, under plain re-execution.
\circled{3}~\emph{Outcome classification} records whether each technique \emph{surfaced} a test's
already-known flakiness, that is, observed at least one passing and one failing verdict, or
\emph{missed} it. \circled{4}~\emph{Empirical analysis} interprets these labels through the three
research questions below.

\subsection{Research Questions}
\label{sec:rqs}

Shaker was originally proposed and evaluated on Android and Java
projects~\cite{silva2020shake,9678918}. Our study asks whether its
concurrency-stress detection strategy transfers to Python, whose flakiness
profile differs substantially~\cite{gruber2021empirical}. We investigate three
questions:

\begin{description}
  \item[\textbf{\EQ{1}}] \ShakerRQOne\\
    \emph{Rationale.} Shaker's added value over simply rerunning a test rests on the claim that
    resource stress amplifies concurrency-induced non-determinism. Comparing it against a
    budget-matched re-execution baseline isolates the contribution of the stress mechanism itself.
  \item[\textbf{\EQ{2}}] \ShakerRQTwo\\
    \emph{Rationale.} A detection technique can only surface flakiness that actually manifests during
    the study. Quantifying how many ground-truth flaky tests reproduce at all on our infrastructure
    bounds the recall any technique can achieve and separates a tool's limitations from the intrinsic
    difficulty of observing flakiness across environments.
  \item[\textbf{\EQ{3}}] \ShakerRQThree\\
    \emph{Rationale.} Understanding which tests reproduce, and which do not, characterizes the classes
    of Python flakiness that stress-based detection can and cannot reach, and points to project
    factors that govern reproducibility.
\end{description}

\subsection{Objects of Analysis}
\label{sec:subjects}

We draw our objects of analysis from the dataset of flaky tests published by Gruber
et al.~\cite{gruber2021empirical, gruber_2021_4450435}, the largest empirical study of flakiness in Python. The
authors collected 876,186 test cases from 22,352 open-source PyPI~\cite{pypi} projects, all using
Pytest, and identified 7,571 flaky tests together with metadata characterizing each test's
behavior and the cause of its flakiness.
The dataset was collected using isolated, containerised test re-execution; that methodology 
was later formalised and open-sourced as the FlaPy framework~\cite{gruber2023flapy}, which 
also produced a larger successor dataset of 2,460 flaky tests from 30,000 PyPI projects.

\MyPara{Selection criteria} We retained the tests Gruber et al. confirm flaky under an unchanged
execution order whose metadata marks them neither order-dependent nor infrastructure-related,
yielding 952 candidates across 277 projects. Excluding order-dependent tests is a control, not a
convenience: Neither technique reorders a suite, so such a test could not surface under either and
would enter as a guaranteed miss for both. Our sample therefore contains \emph{no} order-dependent
tests. Within that pool we did \emph{not} filter by root cause: Restricting it to the resource- and
concurrency-induced flakiness Shaker targets would presuppose our answer and would not reflect
practice, where a test's cause is unknown beforehand. The sample spans randomness, network, timing,
and concurrency alike, which lets RQ3 examine which categories the stress mechanism actually
reaches.

\MyPara{Analyzed sample} Not every candidate could be obtained: Our working set comprises 452 of
the 952 candidate tests, drawn from 159 of the 277 projects. From these we retained the tests
that could be checked out, installed, and executed on our infrastructure under both techniques.
The ReRun baseline produced valid observations for 165 tests and Shaker for 137. Some candidates
could not be prepared or run to completion: Resolving heterogeneous dependencies failed for a
subset, and, under Shaker specifically, the injected stress saturated the host badly enough that a
number of tests could not be installed or could not finish their full run (Section~\ref{sec:threats}).
Because RQ1 is a \emph{paired} comparison, our primary analyzed sample is the \textbf{137 tests
executed under both techniques}; all 137 fall within the 165 ReRun tests, so the two techniques are
compared on an identical set. 
Tests present under only one technique are a source of selection bias we discuss in Section~\ref{sec:threats}.
This attrition rate is consistent with an independent replication on the same
ground truth: iPFlakies could correctly set up only 64\% of the projects it
selected from Gruber et al.'s dataset, citing missing dependencies and
Python-version mismatches as the dominant causes~\cite{wang2022ipflakies},
the same causes we report in Section~\ref{sec:threats}.

\subsection{Detection Procedure}
\label{sec:procedure}

\MyPara{Execution model} Shaker detects flakiness by amplifying non-deterministic behavior through controlled resource stress~\cite{silva2020shake}. Given a budget of $N$
stress runs, it executes the target test $N{\times}4$ times, cycling through four
\texttt{stress-ng}~\cite{stressng} workloads that exert pressure on the CPU and memory subsystems.
The rule by which these runs yield a flakiness verdict is defined in Section~\ref{sec:metrics}.

\MyPara{Budget-matched comparison} To isolate the contribution of resource stress (RQ1), we compare
two techniques executed at an \emph{equal budget of 100 executions per test} under the same
detection criterion, so that any difference is attributable to the stress mechanism rather than to
the number of observations:
\begin{itemize}[leftmargin=1.4em,itemsep=1pt,topsep=2pt]
  \item \textbf{ReRun} performs 100 plain Pytest re-executions with no injected stress
       
  \item \textbf{Shaker} performs 25 stress runs, i.e.\ $25{\times}4 = 100$
        executions, each under one of the four \texttt{stress-ng} workloads.
\end{itemize}
Both techniques thus observe each test 100 times, issued through the same runner
(Listing~\ref{lst:shaker-format}); they differ only in whether those observations are
made under injected resource contention. We set $\text{nsr}=0$ so that all 100 of Shaker's
observations are stressed. Because every object is already a confirmed flaky test, the question is
whether stressed execution surfaces a verdict flip, so spending the entire budget on stressed runs
is the faithful counterpart to ReRun's 100 unstressed runs and keeps the two techniques on an equal
footing.

\MyPara{Unit of evaluation} The ground-truth dataset identifies flakiness at the granularity of an
\emph{individual} test, whereas Shaker was designed to stress an entire suite. Re-running a full
suite for each target test would be prohibitively expensive and would confound the target test's
behavior with that of unrelated tests sharing the run. We therefore evaluate on individual tests in
isolation; doing so required extending Shaker to accept a single test target rather than only a
project directory. Because the projects in our sample declare their dependencies in heterogeneous
formats, we also generalized Shaker's dependency resolution so that each test could be installed and
executed; tests whose dependencies could not be resolved were excluded
(Section~\ref{sec:threats}).

\MyPara{On the execution budget} Our 100-execution budget sits below the ${\sim}170$ re-executions
\citeauthor{gruber2021empirical}~\cite{gruber2021empirical} estimate for 95\% confidence that a
passing test is not flaky; it was chosen so that both techniques could be run on every test at a
feasible cost. The effect of larger budgets is left to future work
(Section~\ref{sec:conclusions}).

\MyPara{Stress workloads} Each stress run cycles the target test through the four
\texttt{stress-ng} workloads of Listing~\ref{lst:stressng}, which pair CPU workers at a
prescribed load with a virtual-memory worker sized as a percentage of available memory. The
values are inherited \emph{unchanged} from the original Shaker
study~\cite{silva2020shake}, where they were tuned for Java and Android; they have never been
calibrated for Python's execution model. Because each test container is capped at 2 CPUs and
2\,GB of memory (Section~\ref{sec:protocol}), these workers contend directly with the test
process. Running the published configuration is what makes the transfer question testable, but
a tuning mismatch may depress detection independently of any fundamental limitation
(Section~\ref{sec:threats}).

\begin{lstlisting}[style=cmd, caption={The four \texttt{stress-ng} workloads, inherited unchanged from the original Shaker study~\cite{silva2020shake}. Each stress run executes the test once under each.}, label={lst:stressng}]
stress-ng --cpu 2 --cpu-load  2 --vm 1 --vm-bytes 21%
stress-ng --cpu 1 --cpu-load 58 --vm 1 --vm-bytes 33%
stress-ng --cpu 1 --cpu-load 61 --vm 1 --vm-bytes 73%
stress-ng --cpu 1 --cpu-load 34 --vm 1 --vm-bytes 37%
\end{lstlisting}

\subsection{Metrics}
\label{sec:metrics}

\MyPara{Detection criterion} A test is \emph{detected} (labelled flaky) when it yields at least one
passing and at least one failing verdict across its runs, and \emph{missed} otherwise. This
criterion is identical to the one \citeauthor{gruber2021empirical}~\cite{gruber2021empirical} use to
establish the ground truth, and is applied identically to Shaker and to the ReRun baseline.

\MyPara{Detection rate} Our primary measure is the \emph{detection rate}: the fraction of the
analyzed ground-truth flaky tests that a technique detects. Because every object is a known flaky
test, this rate is equivalently the technique's \emph{recall} against the ground truth; there are no
true-negative objects, so we do not report precision. We report proportions with 95\% Wilson
confidence intervals~\cite{wilson1927probable}.

\MyPara{Per-question measures} Each research question is answered with a measure derived from the
detection outcome. \emph{RQ1} treats detection as a paired binary outcome per test and compares
Shaker with ReRun using a contingency table (tests detected by both, by Shaker only, by ReRun only,
or by neither) and McNemar's test for paired proportions~\cite{mcnemar1947note}, with a sensitivity analysis excluding
tests with truncated runs. \emph{RQ2} reports the recall of each technique and of their union, and
classifies the non-reproduced tests by the outcome that prevented a verdict. \emph{RQ3} partitions
the sample into reproduced and non-reproduced tests and compares their characteristics, using the
Mann--Whitney~U test with the Cliff's~$\delta$ effect size for numeric attributes and a descriptive
breakdown for the categorical flakiness labels. All statistical tests use a significance level of
$\alpha = 0.05$.

\subsection{Execution Protocol}
\label{sec:protocol}

\MyPara{Isolated, reproducible environments} Each test was executed in a fresh, containerised
environment rebuilt from a fixed software baseline (Python~3.10, the lowest version common to all
tests, and Pytest~7.1.2). For every test, the environment checked out the exact project commit at
which the original flakiness was observed, installed the project's declared dependencies, and ran
the test under Shaker or under plain re-execution. Rebuilding the environment per test prevents
state or dependency leakage between tests and makes each observation independently reproducible.

\MyPara{Execution host} All runs reported here were executed on a single workstation, in Docker
containers with bounded concurrency, each capped at 2 CPUs and 2\,GB of memory, to limit
contention from co-located runs. This differs from the
infrastructure on which the ground truth was originally collected; because flakiness, and injected
stress in particular, interacts with the underlying hardware, this environment difference is a
threat we revisit in Section~\ref{sec:threats} and a phenomenon RQ2 measures directly.

%% file: sections/06_evaluation.tex
\section{Results}
\label{sec:results}

We report our results by research question, combining the detection outcomes of
Shaker and the ReRun baseline with the ground-truth metadata of Gruber
et al.~\cite{gruber2021empirical} (stage~\circled{4} of
Figure~\ref{fig:overview}).

\subsection{RQ1: Shaker vs.\ ReRun}
\label{sec:rq1}

\MyPara{Approach} We compare Shaker against the budget-matched ReRun baseline
(Section~\ref{sec:procedure}) on the \textbf{137 tests executed under both
techniques}, applying the same detection criterion
(Section~\ref{sec:metrics}). Detection is a paired binary outcome per test, so we
cross-tabulate the two techniques (Table~\ref{tab:rq1_contingency}) and test the
difference in proportions with McNemar's test~\cite{mcnemar1947note}.

\input{generated/tab_rq1_contingency}

\begin{figure}
  \centering
  \includegraphics[width=0.82\linewidth]{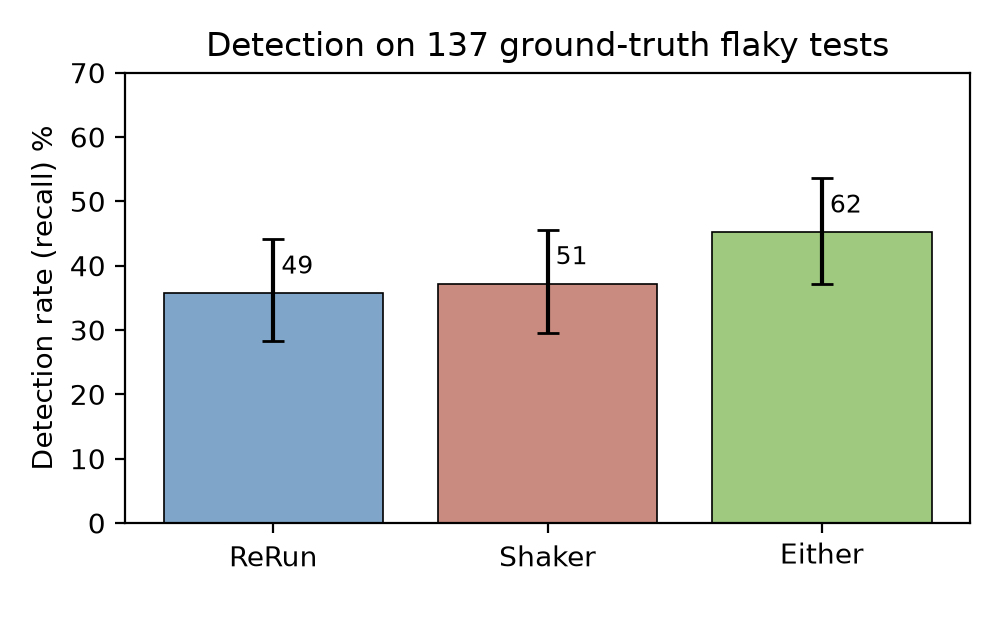}
  \caption{Detection rate (recall) of ReRun, Shaker, and their union
  on the 137 ground-truth flaky tests at a 100-execution budget. Error bars are
  95\% Wilson confidence intervals~\cite{wilson1927probable}; annotations give detected counts.}
  \label{fig:detection}
\end{figure}

\MyPara{Results} Shaker detected flakiness in 51 of the 137 tests (37.2\%,
95\% CI [29.6, 45.6]) and ReRun in 49 (35.8\%, 95\% CI [28.2, 44.1]). Because both
techniques operate on the same tests at the same budget, this comparison
isolates the stress mechanism regardless of the absolute recall level. The two
techniques agree on the large majority of tests: They jointly detect 38 tests
and jointly miss 75, disagreeing on only 24. Of those, Shaker uniquely detects
13 and ReRun uniquely detects 11, a near-symmetric split rather than the
one-sided advantage stress injection would predict. McNemar's test finds no
significant difference (exact two-sided $p = 0.84$). The absolute gap is two
tests, or 1.4 percentage points, narrower than the width of either confidence
interval; with only 24 discordant pairs, any true effect that this test could
have overlooked would itself be small. The conclusion is also robust to
truncated runs: Excluding the 18 tests whose Shaker runs did not complete
(Section~\ref{sec:threats}), a truncation that if anything handicaps Shaker,
leaves the two techniques tied ($p = 1.00$). Figure~\ref{fig:detection} shows
the detection rates with confidence intervals, and
Table~\ref{tab:rq1_contingency} the paired outcomes.

\MyPara{A low joint ceiling} What the paired data expose most sharply is not the
absent gap between the techniques but how much both leave undetected. Their
union detects only 62 of the 137 tests (45.3\%); the remaining 75 (54.7\%)
survive 100 executions under \emph{both} techniques without failing once. The
flat McNemar result shows the shortfall is not something stress addresses: The
tests the two miss, they miss together. The bottleneck at this budget is that
most Python flaky tests fail through mechanisms that neither a bare loop nor CPU
or memory pressure reliably provokes, the categories that dominate RQ3.

\begin{finding}
\textbf{RQ1.} At an equal execution budget, Shaker's stress injection
does \emph{not} detect significantly more flaky Python tests than plain
re-execution (37.2\% vs.\ 35.8\%; McNemar exact $p = 0.84$). The stress
mechanism confers no measurable advantage in this setting.
\end{finding}

\subsection{Case Studies: When Stress Helps and When It Does Not}
\label{sec:cases}

The 24 discordant tests (13 detected only by Shaker and 11 only by ReRun) are,
as a group, statistically indistinguishable from chance (the McNemar result
above). Inspecting individual tests explains \emph{why} the two techniques trade
places rather than one dominating, and confirms that Shaker's mechanism does work
when there is something for it to exploit.

\MyPara{Stress helps: Timing and concurrency} The clearest wins for Shaker are
tests whose flakiness is sensitive to resource pressure or scheduling. The
\texttt{cvbase} timer test of Section~\ref{sec:motivation} is one
(6 failures under stress, 0 without). A stronger case is
\textit{test\_get\_requests\_multiple} from \texttt{selenium-wire}~\cite{seleniumwire_repo}
(Listing~\ref{lst:selwire}), which issues two requests through a recording proxy
and asserts that both were captured. The proxy records requests asynchronously,
so under CPU contention the assertion frequently runs before the second request
is recorded: Shaker failed this test 51 of 100 times, whereas plain re-execution
never failed it. Here stress amplifies a genuine race, exactly as intended.

\begin{lstlisting}[language=Python, caption={A concurrency-sensitive test from \texttt{selenium-wire}~\cite{seleniumwire_repo}.}, label={lst:selwire}]
def test_get_requests_multiple(self):
    self._make_request('https://github.com/')
    self._make_request('https://www.wikipedia.org/')
    # proxy must have recorded both requests by now
    self.assertEqual(2, len(self.client.get_requests()))
\end{lstlisting}

\MyPara{Stress does not help: Randomness} The other side of the split is tests
whose non-determinism has nothing to do with resources. Consider
\textit{test\_run\_d40\_rastrigin} from \texttt{crfmnes}~\cite{crfmnes_repo}
(Listing~\ref{lst:crfmnes}). It asks a \emph{randomized, unseeded} optimizer
(CR-FM-NES) to minimize the 40-dimensional Rastrigin function, a standard
benchmark riddled with local minima, and asserts that the
best value found is essentially the global optimum ($f\_best < 10^{-12}$) within
a fixed iteration budget, an assertion-bound failure mode that recurs across
numerical and machine-learning tests~\cite{dutta2021flex}. Because the optimizer starts from random samples and is
never seeded, each run explores the space of possibilities differently; on an unlucky run it
becomes trapped in a local minimum and fails to reach the optimum in time, so the
assertion fails. This failure probability is a property of the random draws,
not of CPU or memory pressure, so stress cannot change it. Algorithmic
nondeterminism of this kind requires seed- or distribution-aware
techniques~\cite{dutta2020flash} rather than resource stress. The data
confirm this: Both techniques detect the test at similar rates (5 failures under
ReRun, 11 under Shaker, a gap well within run-to-run noise), and the tests ReRun
uniquely caught are of the same randomness-driven character. On this class stress
neither helps nor systematically hurts; it is simply irrelevant.

\begin{lstlisting}[language=Python, caption={A randomness-driven flaky test from \texttt{crfmnes}~\cite{crfmnes_repo}.}, label={lst:crfmnes}]
def test_run_d40_rastrigin():
    # CR-FM-NES: a randomized, *unseeded* optimizer.
    # Rastrigin (dim=40): many local minima to get trapped in.
    _, f_best = CRFMNES(dim=40, f=rastrigin, mean=start,
                        sigma=2.0).optimize(n_iterations)
    # passes only if the random search happens to reach the
    # global optimum (f == 0) within 1e-12 in the given budget
    assert f_best < 1e-12
\end{lstlisting}

These case studies show that Shaker's advantage is real but \emph{narrow}: It
materialises on the timing- and concurrency-sensitive tests it was designed for,
and vanishes on the tests driven by randomness and network behavior that, as RQ3
shows, dominate Python's flaky-test population. Because the amplifiable class is rare
here, the per-test wins never accumulate into an advantage across the population,
which is precisely the null result of RQ1.

\subsection{RQ2: Reproducibility on Independent Infrastructure}
\label{sec:rq2}

\MyPara{Approach} Because every object is a known flaky test, a technique's
detection rate is its \emph{recall} against the ground truth
(Section~\ref{sec:metrics}). RQ1 shows neither technique dominates; RQ2 asks how
high that ceiling is at all. 
Table~\ref{tab:rq2_repro}'s first two rows restate
RQ1's per-technique counts under this reproducibility framing; the quantity RQ1
does not report is the recall of their \emph{union} (a test is reproduced if
\emph{either} technique observes both a pass and a fail). We classify the tests
that no technique reproduces by the outcome that prevented a verdict.

\MyPara{Results} Even taking the union of both techniques, only 62 of the 137
ground-truth flaky tests (45.3\%, 95\% CI [37.2, 53.6]) reproduced as flaky on our
hardware; the majority, 75 tests (54.7\%), never produced a differing verdict
under either technique. The dominant reason is not tool failure but a genuine
absence of observed non-determinism: 61 of the 75 non-reproduced tests passed on
\emph{every} one of their (up to) 100 executions, 13 received too few executions
because their runs were truncated, and only one failed in every execution under
\emph{both} techniques (broken in our environment). That 61 tests never diverged
once across 100 runs indicates a larger budget would recover few of them. In
short, roughly half of the tests labelled flaky by a large-scale study did not
manifest their flakiness when re-executed on different infrastructure,
regardless of whether stress was injected.

\input{generated/tab_rq2_repro}

Our estimate converges with an independent, narrower replication on the same
underlying dataset: using randomized re-execution restricted to order-dependent
tests, iPFlakies reproduced 57\% of previously detected OD
tests~\cite{wang2022ipflakies}, after similarly discarding roughly a third of
candidate projects to dependency and environment failures. Our lower,
category-agnostic figure of 45.3\% is consistent with that estimate once the
denominator is broadened to include network- and randomness-driven tests, which
neither iPFlakies' reordering nor our stress injection targets.

\begin{finding}
\textbf{RQ2.} Fewer than half (45.3\%) of the ground-truth flaky tests reproduce
as flaky on independent hardware under either technique, and most
non-reproductions are tests that never once diverged. Cross-environment
reproducibility, not the choice of detection mechanism, is the dominant limiting
factor.
\end{finding}

\subsection{RQ3: Characteristics of Reproduced vs.\ Missed Tests}
\label{sec:rq3}

\MyPara{Approach} We join each test's outcome (reproduced by either technique
vs.\ not) with two sources of metadata. The first comprises the manual
root-cause flakiness categories from the \texttt{100NOD} sample, a manually classified
set of 100 \emph{non-order-dependent} (NOD, i.e.\ flaky regardless of test
execution order) flaky tests from Gruber et al.~\cite{gruber2021empirical, gruber_2021_4450435},
available for the subset of tests it labels. The second comprises
lightweight project traits (size in LOC, number of files, test files, and
declared dependencies) mined from a shallow checkout of each project at its
recorded commit. We compare numeric traits between the reproduced and
non-reproduced groups with the Mann--Whitney~U test and Cliff's~$\delta$ effect
size, and summarise the categorical breakdown descriptively.

\MyPara{Results} Two patterns emerge (see Table~\ref{tab:rq3_traits}). First, the
flakiness in this sample is dominated by causes other than concurrency: Across
the manual classification, network and randomness account for the bulk of
labelled tests, while the only concurrency-adjacent category present,
async wait, comprises just three tests. Among the labelled tests,
flakiness caused by randomness reproduced readily (13/19) while flakiness caused
by network interactions almost never did (1/6); neither is a class that resource
stress is expected to surface. Within this labelled subset ($n=31$ of the 137
analyzed tests), this pattern is consistent with the null RQ1 result: The
stress mechanism has little concurrency non-determinism to amplify. We caution
that the labelled subset, being manually classified, may not be representative 
of the full sample's cause distribution.
Second, reproduced tests come from significantly \emph{smaller} projects than 
missed tests (median 1{,}041 vs.\ 3{,}551 LOC; Mann--Whitney $p < 0.001$,
Cliff's $\delta = -0.36$, a medium effect), consistent with larger projects 
being harder to reproduce faithfully on independent infrastructure. 
The remaining structural traits (numbers of files, test files, and 
dependencies) show no significant difference.

\input{generated/tab_rq3_traits}

\begin{finding}
\textbf{RQ3.} The reproducible flaky tests are dominated by network and
randomness rather than concurrency, and cluster in smaller projects (medium
effect). Both observations explain why stress injection adds no value here: The
flakiness present is largely outside the mechanism's scope.
\end{finding}

\subsection{Discussion}
\label{sec:discussion}

Taken together, the three findings tell a coherent story about transferring a
Java/Android stress-based detector to Python. Stress injection does not beat
plain re-execution (RQ1) not because the tool is broken, but because the
flakiness available to detect is largely the wrong \emph{kind}: On independent
hardware fewer than half of the ground-truth tests reproduce at all (RQ2), and
those that do are overwhelmingly driven by network behavior and randomness
rather than concurrency (RQ3). 

\MyPara{Practical implications} For practitioners, adopting Shaker on Python offers no detection benefit over
simply rerunning tests at the same budget. It also adds an operational cost,
provisioning \texttt{stress-ng} and tolerating the truncated runs a saturated
host produces, that a CI pipeline would pay on every invocation. Teams
choosing between the two techniques for Python test suites can therefore default
to plain re-execution without a measurable loss in detection power.

\MyPara{A hazard for the field: Flaky-test ground truths do not travel}
For researchers, the
reproducibility gap is a cautionary result in its own right, and one that outlives the particular
tool we evaluated. Reusing a flaky-test ground truth collected on one infrastructure to evaluate a
tool on another silently converts genuine flaky tests into apparent true negatives, deflating any
tool's measured recall regardless of the technique under study. Here that deflation is not a
rounding error: More than half of our objects never flipped once. We therefore recommend that
studies reusing such a ground truth report the reproduction rate observed in their own environment
alongside any recall figure, and read recall measured against an external ground truth as a
\emph{lower bound} rather than an estimate.
Whether stress configurations \emph{calibrated} for Python's execution model
could recover the concurrency cases, and how detection scales with larger
execution budgets, remain open questions we could not answer with the available
data (Section~\ref{sec:conclusions}).

%% file: generated/tab_rq1_contingency.tex
\begin{table}
\centering
\begin{tabular}{l c c c}
\toprule
 & \textbf{ReRun detects} & \textbf{ReRun misses} & \textbf{Total} \\
\midrule
\textbf{Shaker detects} & 38 & 13 & 51 \\
\textbf{Shaker misses}  & 11 & 75 & 86 \\
\midrule
\textbf{Total}          & 49 & 88 & 137 \\
\bottomrule
\end{tabular}
\caption{Paired detection outcomes for Shaker versus the budget-matched ReRun
baseline on the 137 ground-truth flaky tests, 100 executions each. The
difference is not significant (McNemar exact $p = 0.84$).}
\label{tab:rq1_contingency}
\end{table}

%% file: generated/tab_rq2_repro.tex
\begin{table}
\centering
\begin{tabular}{l c c}
\toprule
\textbf{Technique / outcome} & \textbf{Count} & \textbf{Rate} \\
\midrule
Reproduced by Shaker (25 stress runs) & 51/137 & 37.2\% \\
Reproduced by ReRun (100 re-executions) & 49/137 & 35.8\% \\
Reproduced by \emph{either} technique & 62/137 & 45.3\% \\
\midrule
\multicolumn{3}{l}{\emph{Not reproduced by either (75/137), by reason:}} \\
\quad all\_passed (never flipped) & 61 & 81.3\% \\
\quad truncated (under-sampled) & 13 & 17.3\% \\
\quad all\_failed (broken in env) & 1 & 1.3\% \\
\bottomrule
\end{tabular}
\caption{Reproducibility of the 137 ground-truth flaky tests on independent
hardware, with the reason each non-reproduced test yielded no verdict flip.}
\label{tab:rq2_repro}
\end{table}

%% file: generated/tab_rq3_traits.tex
\begin{table}[h]
\centering
\small
\textbf{(a) Reproduction by manual flakiness category} ($n=31$ labelled)\\[2pt]
\begin{tabular}{l c c c}
\toprule
\textbf{Category} & \textbf{Tests} & \textbf{Reproduced} & \textbf{Rate} \\
\midrule
random & 19 & 13 & 68\% \\
network & 6 & 1 & 17\% \\
async wait & 3 & 2 & 67\% \\
IO & 1 & 0 & 0\% \\
resource leak & 1 & 1 & 100\% \\
time & 1 & 0 & 0\% \\
\bottomrule
\end{tabular}

\vspace{6pt}
\textbf{(b) Project traits: reproduced vs.\ not} ($n=137$ with metadata)\\[2pt]
\begin{tabular}{l c c c c}
\toprule
\textbf{Trait} & \textbf{Med.\ repro} & \textbf{Med.\ not} & \textbf{$p$} & \textbf{Cliff's $\delta$} \\
\midrule
Project size (LOC) & 1041 & 3551 & 0.00 & -0.36 (medium) \\
\# files & 26 & 39 & 0.06 & -0.19 (small) \\
\# test files & 4 & 6 & 0.53 & -0.06 (negligible) \\
\# dependencies & 4 & 4 & 0.43 & -0.08 (negligible) \\
\bottomrule
\end{tabular}
\caption{Reproduced versus non-reproduced flaky tests: (a)~reproduction by
manual flakiness category; (b)~project traits compared with the Mann--Whitney U
test and Cliff's~$\delta$.}
\label{tab:rq3_traits}
\end{table}

%% file: sections/07_threats.tex
\section{Threats to Validity}
\label{sec:threats}

We discuss threats to validity under the standard internal, external, and construct
classification~\cite{wohlin2012experimentation}. Because this study repurposes an existing tool and
an existing ground truth for a new ecosystem and a new execution environment, most of them stem
from that adaptation.

\subsection{Internal Validity}

\MyPara{Cross-environment reproducibility} The ground truth of Gruber
et al.~\cite{gruber2021empirical} was collected on different infrastructure from ours. Because
flakiness is by definition environment-sensitive, and its manifestation depends on runtime
conditions such as timing and scheduling~\cite{lam2019rootcausing}, tests labelled flaky there may
not manifest flakiness here. 
This gap was also independently reported for the order-dependent subset of the same ground 
truth by iPFlakies~\cite{wang2022ipflakies}.
Rather than treat this only as a nuisance, we measure it directly: RQ2 quantifies
how many tests reproduce (45.3\% under either technique) and shows that most non-reproductions are
tests that never diverged. This is the central internal-validity caveat of the study, and it applies
symmetrically to Shaker and to ReRun, so it does not bias the RQ1 comparison between them.

\MyPara{Truncated runs} 18 of the 137 tests did not complete the full 100 executions under Shaker,
against only 3 under ReRun. The asymmetry is itself informative: The injected stress saturates the
single host, so runs are more likely to be killed under Shaker than under plain re-execution.
Truncation can only lower the chance of observing a verdict flip, so it works \emph{against} Shaker;
the null RQ1 result is therefore conservative toward the technique. To confirm that truncation does
not drive the conclusion, we repeated the comparison on the 119 tests with complete runs, and Shaker
and ReRun remain statistically indistinguishable ($p = 1.00$).

\MyPara{Single-test execution} The ground truth was collected by running whole suites; we execute
each target test in isolation (Section~\ref{sec:procedure}). Isolation removes the intra-suite
contention and shared state that were part of the conditions under which the original flakiness
appeared, making it a plausible co-explanation for the reproducibility gap RQ2 measures: A test
whose flakiness depends on its suite-mates cannot flip when run alone, however often it is
repeated. The constraint applies identically to both techniques, so it does not bias RQ1;
suite-granularity replication is left to future work (Section~\ref{sec:conclusions}).

\MyPara{Selection bias from setup failures} Not every candidate test could be checked out, installed,
and executed on our infrastructure: The ReRun baseline yielded valid observations for 165 tests and
Shaker for 137, and our paired analysis is restricted to the 137 tests common to both. 
If the tests that failed to set up differ systematically in their flakiness characteristics from those that ran,
our sample is not fully representative. We report the coverage gap explicitly rather than silently
dropping tests.

\MyPara{Single-host execution} All runs reported here were executed on one workstation. This removes
the machine-to-machine confound that pooling heterogeneous hosts would introduce, but it also means
our reproducibility figures are specific to one environment and should not be read as a universal
reproduction rate.

\subsection{External Validity}

\MyPara{Sample size and scope} Our paired analysis covers 137 tests on a single infrastructure. This
is a focused replication, not a large-scale study, and its quantitative estimates carry
correspondingly wide confidence intervals (reported throughout Section~\ref{sec:results}). We frame
our conclusions as evidence about whether stress-based detection \emph{transfers} to Python, not as
precise population estimates.

\MyPara{Order-dependent flakiness is out of scope} Our sample excludes order-dependent tests by
construction, since neither technique reorders a suite (Section~\ref{sec:subjects}). The exclusion
is necessary for a fair comparison but bounds what our results say: Order dependency is Python's
largest flakiness category, 59\% of the flaky tests in Gruber et al.'s ground truth, and our
findings speak to none of it. Reordering tools such as iDFlakies~\cite{lam2019idflakies} and
iPFlakies~\cite{wang2022ipflakies} are the appropriate instruments there.

\MyPara{Dataset and version constraints} The dataset is limited to open-source PyPI projects that use
Pytest and were active at the time of \citeauthor{gruber2021empirical}'s study, executed under a fixed baseline
(Python~3.10, Pytest~7.1.2). Results may not generalize to proprietary projects, other frameworks, or
newer language and framework versions. The two Shaker modifications (\texttt{-{}-specific-tests-path},
generalized dependency handling) are specific to this study's setup and may require adaptation for
arbitrary Python projects.

\subsection{Construct Validity}

\MyPara{Detection criterion} A test is declared flaky if it produces at least one passing and one
failing verdict across its executions. Using the same criterion for the ground truth and for both
detection techniques is a deliberate methodological choice that makes the comparison valid and places
Shaker and ReRun on equal terms. The criterion cannot, however, distinguish flakiness triggered by
stress injection from flakiness that would occur regardless of stress.

\MyPara{What the budget equalizes} Matching the techniques at 100 executions equalizes the number of
\emph{observations}, the confound RQ1 must control, but not their cost: Shaker keeps
\texttt{stress-ng} workers in the container throughout, so at an identical observation count it
consumes strictly more CPU, memory, and wall-clock time. The 18 truncated Shaker runs against 3
under ReRun are that cost becoming visible. The asymmetry works \emph{against} Shaker on cost while
leaving its detection rate unchanged, so it strengthens our conclusion.

\MyPara{Budget allocation within Shaker} We ran Shaker with $\text{nsr}=0$, spending the entire
budget on stressed runs as the faithful counterpart to ReRun's 100 unstressed runs
(Section~\ref{sec:procedure}). One consequence is that Shaker's mode that splits the budget between
stressed and unstressed runs was not exercised; such a split might behave differently and is worth
exploring in future work.

\MyPara{Uncalibrated stress configurations} The \texttt{stress-ng} configurations were taken directly
from the original Shaker paper~\cite{silva2020shake}, where they were tuned for Java/Android. They
have never been calibrated for Python. If they are inappropriate for Python's execution model, Shaker
may underperform because of a tuning mismatch rather than a fundamental limitation, a possibility our
data cannot rule out and that we flag as future work (Section~\ref{sec:conclusions}).

%% file: sections/08_related_work.tex
\section{Related Work}
\label{sec:related_work}

\MyPara{Empirical studies of flaky tests} \citeauthor{luo2014empirical}~\cite{luo2014empirical} conducted
the foundational empirical study of flaky tests, analyzing 201 commits from 51
open-source Java projects and establishing a taxonomy of ten root cause categories,
with asynchronous waits, concurrency, and order dependency as the most prevalent.
\citeauthor{parry2021survey}~\cite{parry2021survey} synthesized 76 papers on flaky tests across four
dimensions (causes, costs, detection, and mitigation), providing a broad survey of the
field. \citeauthor{eck2019understanding}~\cite{eck2019understanding} investigated the
developer perspective through a survey of 121 practitioners, finding that reproducing
flakiness and diagnosing its root cause are the most challenging aspects. 
\citeauthor{gruber2021empirical}~\cite{gruber2021empirical} conducted the first large-scale study specifically for
Python, analyzing 876,186 test cases across 22,352 open-source projects and finding that
order dependency (59\%) and infrastructure (28\%) dominate Python flakiness, a
distribution markedly different from Java, where concurrency plays a larger relative role.
Our work is the first to evaluate Shaker, a stress-based detection tool for
concurrency-related flakiness, in this Python-specific context.

\MyPara{Flaky test detection tools} Several tools have been proposed to detect flaky
tests more efficiently than naive re-execution. DeFlaker~\cite{bell2018deflaker}
detects flaky tests without rerunning by monitoring code coverage changes between test
runs: a test is flagged as potentially flaky if it fails while covering no code changed
since the last passing run. Evaluated on 96 Java projects, DeFlaker detected 1,874 flaky
tests out of 4,846 failures, with a 1.5\% false-alarm rate. 
iDFlakies~\cite{lam2019idflakies}
detects order-dependent and non-order-dependent flaky tests by running the test suite in
randomized orders. Its curated dataset of 422 confirmed flaky tests has become a benchmark
for subsequent work. 
iPFlakies~\cite{wang2022ipflakies} extends this reordering strategy to
Python, and is the closest prior work to ours in that it too re-executes a subset of Gruber
et al.'s ground truth on independent infrastructure, reporting a similar reproducibility gap
for order-dependent tests. 
Shaker~\cite{silva2020shake, 9678918} is orthogonal to all three: instead of reordering tests
or tracking coverage, it \textit{amplifies} concurrency-induced non-determinism by
injecting CPU, memory, and I/O stress during test execution. Unlike DeFlaker and iDFlakies,
Shaker specifically targets concurrency-related flakiness and requires no code
instrumentation. Predictive approaches such as FlakeFlagger~\cite{alshammari2021flakeflagger}
instead avoid extra executions altogether, classifying tests as flaky from behavioural features or
from source-code vocabulary~\cite{pinto2020vocabulary}.
Re-execution (ReRun) remains the de facto baseline in practice and is the point of comparison we
adopt for Shaker.

\MyPara{Shaker} \citeauthor{silva2020shake}~\cite{silva2020shake} introduced Shaker and 
evaluated it on a benchmark of 75 flaky tests from 11 Android applications. 
This evaluation reported a substantial, consistent advantage over ReRun in detection rates and revealing novel flaky tests.
A subsequent publication~\cite{9678918} described Shaker's GitHub Action integration and CLI modes.
Our work is the first evaluation of Shaker in the Python ecosystem: at an equal execution 
budget we observe 37.2\% for Shaker against 35.8\% for ReRun, a difference that is not 
statistically significant, in sharp contrast to the advantage Shaker showed on Java and
Android.

%% file: sections/09_conclusions.tex
\section{Conclusions}
\label{sec:conclusions}

Shaker accelerates flaky-test detection by injecting resource stress, and was reported to outperform
plain re-execution on Java and Android. This paper presents its first empirical evaluation for
Python, using ground-truth flaky tests from the dataset of
\citeauthor{gruber2021empirical}~\cite{gruber2021empirical}. Across a paired comparison on 137 ground-truth flaky tests, stress injection provided no
statistically significant detection advantage over plain re-execution (RQ1: 37.2\% vs.\ 35.8\%,
$p = 0.84$), because fewer than half of the ground-truth tests reproduce on independent hardware at
all (RQ2: 45.3\%), and the tests that do reproduce are dominated by network and randomness rather
than the concurrency Shaker targets (RQ3). The tool itself is not at fault: The flakiness available
to detect lies largely outside its mechanism's reach. The most consequential finding, however, is
the one that outlives Shaker. A flaky-test ground truth does not travel between execution
environments, so any recall measured against a ground truth collected elsewhere is a lower bound of
unknown slack, and cross-paper comparisons of detection rates rest on weaker ground than the field
usually assumes (Section~\ref{sec:discussion}).

Three questions our data could not settle remain open: whether detection improves materially as the
execution budget grows past the $\sim$170-rerun threshold of \citeauthor{gruber2021empirical};
whether \texttt{stress-ng} configurations tuned for Python's execution model, rather than inherited
from Java and Android, recover the concurrency cases the default configuration misses; and whether
running each test within its suite, rather than in isolation, restores part of the flakiness that
did not reproduce here. Answering any of them would sharpen our understanding of where, if anywhere,
stress injection pays off in Python.

%% file: biblio.bib
@inproceedings{gruber2023flapy,
author = {Gruber, Martin and Fraser, Gordon},
title = {{FlaPy}: Mining Flaky {P}ython Tests at Scale},
year = {2023},
publisher = {IEEE Press},
booktitle = {45th International Conference on Software Engineering: Companion Proceedings},
}

@INPROCEEDINGS{wang2022ipflakies,
  author={Wang, Ruixin and Chen, Yang and Lam, Wing},
  booktitle={44th International Conference on Software Engineering: Companion Proceedings}, 
  title={iPFlakies: A Framework for Detecting and Fixing Python Order-Dependent Flaky Tests}, 
  year={2022},
 }

@article{mcnemar1947note,
  title={Note on the sampling error of the difference between correlated proportions or percentages},
  author={McNemar, Quinn},
  journal={Psychometrika},
  volume={12},
  number={2},
  pages={153--157},
  year={1947},
  publisher={Springer-Verlag}
}

@article{wilson1927probable,
  title={Probable inference, the law of succession, and statistical inference},
  author={Wilson, Edwin B},
  journal={Journal of the American Statistical Association},
  volume={22},
  number={158},
  pages={209--212},
  year={1927},
  publisher={Taylor \& Francis}
}

@article{parry2021survey,
  title={A Survey of Flaky Tests},
  author={Parry, Owain and Kapfhammer, Gregory M. and Hilton, Michael and McMinn, Phil},
  journal={ACM Transactions on Software Engineering and Methodology},
  year={2021},
}

@inproceedings{eck2019understanding,
  title={Understanding Flaky Tests: The Developer's Perspective},
  author={Eck, Moritz and Palomba, Fabio and Castelluccio, Marco and Bacchelli, Alberto},
  booktitle={Proceedings of the 27th ACM Joint European Software Engineering Conference and
             Symposium on the Foundations of Software Engineering (ESEC/FSE)},
  year={2019},
}

@inproceedings{bell2018deflaker,
  title={{DeFlaker}: Automatically Detecting Flaky Tests},
  author={Bell, Jonathan and Legunsen, Owolabi and Hilton, Michael and Eloussi, Lamyaa
          and Yung, Tifany and Marinov, Darko},
  booktitle={Proceedings of the 40th International Conference on Software Engineering (ICSE)},
  year={2018},
}

@inproceedings{lam2019idflakies,
  title={iDFlakies: A framework for detecting and partially classifying flaky tests},
  author={Lam, Wing and Oei, Reed and Shi, August and Marinov, Darko and Xie, Tao},
  booktitle={2019 12th ieee conference on software testing, validation and verification (icst)},
  pages={312--322},
  year={2019},
  organization={IEEE}
}

@inproceedings{silva2020shake,
  title={Shake it! detecting flaky tests caused by concurrency with shaker},
  author={Silva, Denini and Teixeira, Leopoldo and d'Amorim, Marcelo},
  booktitle={2020 IEEE International Conference on Software Maintenance and Evolution (ICSME)},
  pages={301--311},
  year={2020},
  organization={IEEE}
}

@inproceedings{9678918,
  author={Cordeiro, Marcello and Silva, Denini and Teixeira, Leopoldo and Miranda, Breno and d'Amorim, Marcelo},
  booktitle={36th IEEE/ACM International Conference on Automated Software Engineering (ASE)},
  title={Shaker: a Tool for Detecting More Flaky Tests Faster},
  year={2021},
 }

@inproceedings{gruber2021empirical,
    title={An empirical study of flaky tests in python},
    author={Gruber, Martin and Lukasczyk, Stephan and Kroi{\ss}, Florian and Fraser, Gordon},
    booktitle={2021 14th IEEE Conference on Software Testing, Verification and Validation (ICST)},
    pages={148--158},
    year={2021},
    month = {apr},
    organization={IEEE}
}

@dataset{gruber_2021_4450435,
    author = {Gruber, Martin},
    title = {Dataset of An Empirical Study of Flaky Tests in Python},
    month = jan,
    year = 2021,
    publisher = {Zenodo},
    doi = {10.5281/zenodo.4450435},
    url = {https://doi.org/10.5281/zenodo.4450435}
}

@inproceedings{luo2014empirical,
  title={An empirical analysis of flaky tests},
  author={Luo, Qingzhou and Hariri, Farah and Eloussi, Lamyaa and Marinov, Darko},
  booktitle={Proceedings of the 22nd ACM SIGSOFT international symposium on foundations of software engineering},
  pages={643--653},
  year={2014}
}

@misc{Google,
  title = {Flaky Tests at Google and How We Mitigate Them},
  author={John Micco},
  year={2016},
  howpublished = {\url{https://testing.googleblog.com/2016/05/flaky-tests-at-google-and-how-we.html}}
}

@misc{Pytest,
  title = {Pytest Documentation},
  author = {{Pytest Team}},
  year={2022},
  howpublished = {\url{https://docs.pytest.org/en/7.1.x/}}
}

@misc{ieee,
  title = {Top Programming Languages 2025},
  author = {{IEEE Spectrum}},
  year={2025},
  howpublished = {\url{https://spectrum.ieee.org/top-programming-languages-2025}}
}

@misc{stressng,
  title = {stress-ng},
  author = {King, Colin},
  year={2022},
  howpublished = {\url{https://github.com/ColinIanKing/stress-ng}}
}

@misc{maven,
  title = {Apache Maven},
  author = {{Apache Software Foundation}},
  year={2022},
  howpublished = {\url{https://maven.apache.org/}}
}

@misc{pypi,
  title = {PyPI},
  author = {{Python Software Foundation}},
  year={2022},
  howpublished = {\url{https://pypi.org/}}
}

@article{lam2020longitudinal,
  title={A Large-Scale Longitudinal Study of Flaky Tests},
  author={Lam, Wing and Winter, Stefan and Wei, April and Xie, Tao and Marinov, Darko and Bell, Jonathan},
  journal={Proceedings of the ACM on Programming Languages},
  volume={4},
  number={OOPSLA},
  pages={1--28},
  year={2020},
  publisher={ACM},
  doi={10.1145/3428270}
}

@inproceedings{alshammari2021flakeflagger,
  title={{FlakeFlagger}: Predicting Flakiness Without Rerunning Tests},
  author={Alshammari, Abdulrahman and Morris, Christopher and Hilton, Michael and Bell, Jonathan},
  booktitle={Proceedings of the 43rd International Conference on Software Engineering (ICSE)},
  year={2021},
}

@book{wohlin2012experimentation,
  title={Experimentation in software engineering},
  author={Wohlin, Claes and Runeson, Per and H{\"o}st, Martin and Ohlsson, Magnus C and Regnell, Bj{\"o}rn and Wessl{\'e}n, Anders and others},
  year={2012},
  publisher={Springer}
}

@misc{cvbase_repo,
  author       = {Cvbase},
  year         = {2026},
  title        = {Cvbase},
  howpublished = {\url{https://github.com/hellock/cvbase}},
}

@misc{seleniumwire_repo,
  author       = {Selenium Wire},
  year         = {2026},
  title        = {Selenium-wire},
  howpublished = {\url{https://github.com/wkeeling/selenium-wire}},
}

@misc{crfmnes_repo,
  author       = {CR-FM-NES},
  year         = {2026},
  title        = {Ccrfmnes},
  howpublished = {\url{https://github.com/nomuramasahir0/crfmnes}},
}

@inproceedings{memon2017taming,
  title={Taming Google-scale continuous testing},
  author={Memon, Atif and Gao, Zebao and Nguyen, Bao and Dhanda, Sanjeev and Nickell, Eric and Siemborski, Rob and Micco, John},
  booktitle={2017 IEEE/ACM 39th International Conference on Software Engineering: Software Engineering in Practice Track (ICSE-SEIP)},
  pages={233--242},
  year={2017},
  organization={IEEE}
}

@inproceedings{lam2019rootcausing,
  title={Root causing flaky tests in a large-scale industrial setting},
  author={Lam, Wing and Godefroid, Patrice and Nath, Suman and Santhiar, Anirudh and Thummalapenta, Suresh},
  booktitle={28th ACM SIGSOFT International Symposium on Software Testing and Analysis},
  year={2019}
}

@inproceedings{dutta2020flash,
  title={Detecting flaky tests in probabilistic and machine learning applications},
  author={Dutta, Saikat and Shi, August and Choudhary, Rutvik and Zhang, Zhekun and Jain, Aryaman and Misailovic, Sasa},
  booktitle={Proceedings of the 29th ACM SIGSOFT international symposium on software testing and analysis},
  pages={211--224},
  year={2020}
}

@inproceedings{dutta2021flex,
  title={Flex: fixing flaky tests in machine learning projects by updating assertion bounds},
  author={Dutta, Saikat and Shi, August and Misailovic, Sasa},
  booktitle={Proceedings of the 29th ACM Joint Meeting on European Software Engineering Conference and Symposium on the Foundations of Software Engineering},
  pages={603--614},
  year={2021}
}

@inproceedings{pinto2020vocabulary,
  title={What is the vocabulary of flaky tests?},
  author={Pinto, Gustavo and Miranda, Breno and Dissanayake, Supun and d'Amorim, Marcelo and Treude, Christoph and Bertolino, Antonia},
  booktitle={17th International Conference on Mining Software Repositories},
  year={2020}
}
